\documentclass{article}

\usepackage{arxiv}

\usepackage[utf8]{inputenc} 
\usepackage[T1]{fontenc}    
\usepackage{hyperref}       
\usepackage{url}            
\usepackage{booktabs}       
\usepackage{amsfonts}       
\usepackage{nicefrac}       
\usepackage{microtype}      
\usepackage{lipsum}		
\usepackage{graphicx}
\usepackage{natbib}
\usepackage{doi}
\usepackage{cancel}
\usepackage{amsmath}

\title{Bayesian Surrogate Analysis and Uncertainty Propagation}

\author{ \href{https://orcid.org/0000-0001-8956-2576}{\includegraphics[scale=0.06]{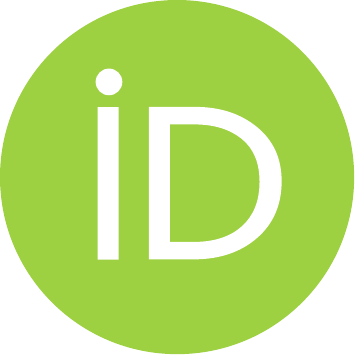}\hspace{1mm}Sascha Ranftl}  \\
	Institute of Theoretical Physics-Computational Physics\\
	Graz University of Technology \\
	8010 Graz, Austria \\
	\texttt{ranftl@tugraz.at} \\
	\And
	\href{https://orcid.org/0000-0001-7436-5078}{\includegraphics[scale=0.06]{orcid.pdf}\hspace{1mm}Wolfgang von der Linden} \\
	Institute of Theoretical Physics-Computational Physics\\
	Graz University of Technology \\
	8010 Graz, Austria \\
}

\renewcommand{\shorttitle}{\textit{arXiv} Template}

\hypersetup{
pdftitle={A template for the arxiv style},
pdfsubject={q-bio.NC, q-bio.QM},
pdfauthor={David S.~Hippocampus, Elias D.~Striatum},
pdfkeywords={First keyword, Second keyword, More},
}

\begin{document}
\maketitle

\begin{abstract}
The quantification of uncertainties of computer simulations due to input parameter uncertainties is paramount to assess a model's credibility. For computationally expensive simulations, this is often feasible only via surrogate models that are learned from a small set of simulation samples. The surrogate models are commonly chosen and deemed trustworthy based on  heuristic measures, and substituted for the simulation in order to approximately propagate the simulation input uncertainties to the simulation output. In the process, the contribution of the uncertainties of the surrogate itself to the simulation output uncertainties are usually neglected.  In this work, we specifically address the case of doubtful surrogate trustworthiness, i.e. non-negligible surrogate uncertainties. 
We find that Bayesian probability theory yields a natural measure of surrogate trustworthiness, and that surrogate uncertainties can easily be included in simulation output uncertainties. For a Gaussian likelihood for the simulation data, with unknown surrogate variance and given a generalized linear surrogate model, the resulting formulas reduce to simple matrix multiplications. The framework contains Polynomial Chaos Expansions as a special case, and is easily extended to Gaussian Process Regression. Additionally, we show a simple way to implicitly include spatio-temporal correlations.
Lastly, we demonstrate a numerical example where surrogate uncertainties are in part negligible and in part non-negligible.
\end{abstract}

\keywords{uncertainty quantification \and uncertainty propagation \and surrogate models \and meta-modelling, Bayesian analysis}

\section{Introduction}

 
%
Uncertainty quantification of simulations has gained increasing attention, e.g. in the field of Computational Engineering, in order to address doubtful parameter choices and assess the models' credibility. Surrogate models have become a popular tool to propagate simulation input uncertainties to the simulation output, particularly for modern day applications with high computational cost and many uncertain model parameters. For that, a parametrized surrogate model (synonyms: meta-model, emulator) is learned from a finite set of simulation samples. I.e. the surrogate is a function of the uncertain simulation input parameters that is 'fitted' to the simulation output data. The quality of this fit is then judged by heuristic diagnostics, and the surrogate deemed trustworthy respectively. A key aspect of this procedure is, that the surrogate can be evaluated much faster than the simulation, and still retains a reasonable approximation to the simulation. 

The simulation is then substituted with the surrogate model in order to compute the marginal probability density function of the simulation output.  The simulation uncertainties are thus inferred from the surrogate model instead of the original simulation model at a significantly reduced computational effort. While this practise allows to obtain estimates on uncertainties of expensive simulations in the first place, the contribution of the uncertainty of the surrogate itself to the total simulation uncertainty is commonly neglected. In other words, the estimation of the surrogate parameters based on the finite set of simulation samples entails an additional uncertainty in the sought-for uncertainty of the simulation output. The purpose of this paper is to investigate this surrogate uncertainty as the natural measure for the surrogate's trustworthiness, and how the surrogate uncertainty affects the simulation output uncertainty. 

In many cases, the surrogate uncertainty is indeed small if the heuristic diagnostics naively imply so. If the heuristic diagnostics imply that the surrogate is not trustworthy, one may resort to two options: (i) Acquire more simulation data until the surrogate is trustworthy. This is limited by the computational budget and the surrogate's convergence properties. (ii)  Shrink the parameter space, e.g. omit a number of uncertain simulation parameters by assuming definite parameter values. However, in some cases (i) is not feasible and  (ii) is not desired. In this contribution, we demonstrate how to include surrogate uncertainties if the user deals with a surrogate model with doubtful trustworthiness. 

Popular surrogate models are Polynomial Chaos Expansions
\citep{Xiu2005, OHagan2013, Crestaux2009} and Gaussian Process Regression  \citep{OHagan1978, Rasmussen2006}, the latter of which has had its renaissance recently from within the machine learning community.  In this work, we assume a Gaussian likelihood for the simulation data with unknown variance and given a generalized linear surrogate model (i.e. linear in the surrogate parameters) that includes Polynomial Chaos Expansions as a special case and is easily extended to Gaussian Process Regression.  
Other Bayesian perspectives on Uncertainty Quantification of computer simulations with these popular surrogate models are given in \citep{Sraj2016,OHagan1999,Kennedy2000,Arnst2010,Madankan2013,Karagiannis2014,Lu2015a,Hwai2015}.
A comprehensive collection of reviews on Uncertainty Quantification, from the point of view of computational engineering and applied mathematics, can be found in \citep{Ghanem2017}. In \citep{OHagan2006, OHagan1999}, a statistician's perspective is discussed. Here, we will use Bayesian Probability Theory \citep{vonderLinden2014}.


\section{Bayesian Uncertainty Quantification} \label{sec:buq}
%
\LetLtxMacro{\originaleqref}{\eqref}
\renewcommand{\eqref}{eq.~\originaleqref}
\newcommand\figref{fig.~\ref}
\newcommand\sectref{sect.~\ref}
\newcommand\tabref{table~\ref}

\newcommand{\expect}[1]{\langle #1  \rangle}
\newcommand{\condexpect}[2]{\langle #1 \mid #2 \rangle}
\newcommand{\BK}{\mathcal{I}}
\newcommand{\p}[2]{{p}\big(#1   \boldsymbol{\mid}  #2 \big)}
\newcommand{\prior}[1]{{p} \big(#1\big)}
\newcommand{\pN}[2]{\mathcal{N}\Big[#1 {\color{red} \boldsymbol{\mid}} #2 \Big]}
\newcommand{\pr}{^{\prime}}
\renewcommand{\a}{\boldsymbol{a}}
\renewcommand{\d}{\boldsymbol{{d}}}
\newcommand{\gp}{\tilde{z}}
\renewcommand{\b}{\boldsymbol{\beta}}
\newcommand{\de}{\d_{exp}}
\newcommand{\ds}{\d_{sim}}
\newcommand{\hparams}{c}
\newcommand{\vhparams}{\boldsymbol{\hparams}}
\newcommand{\z}{z}
\newcommand{\boldz}{\boldsymbol{\z}}
\newcommand{\pq}{\boldsymbol{p}}
\newcommand{\st}{_{\ast}}
\newcommand{\etav}{\boldsymbol{\eta}}
\newcommand{\gamdotv}{\boldsymbol{\dot \gamma}}
\renewcommand{\a}{\boldsymbol{a}}
\newcommand{\dgamma}{\dot{\gamma}}
\newcommand{\x}{\dgamma}
\newcommand{\vv}[1]{{\boldsymbol #1}}
\newcommand{\avg}[1]{\left\langle #1 \right \rangle}
\newcommand{\opttxt}[1]{\textcolor{blue}{#1}}
\newcommand{\soutw}[1]{\sout{\textcolor{white}{#1}}}

\newcommand{\dexp}{\vv d_\text{exp}}
\newcommand{\dsim}{\vv d_\text{sim}}
\newcommand{\Dsim}{\vv D_\text{sim}}

\newcommand{\Wo}{\text{Wo}}
\newcommand{\Rey}{\text{Re}}

\newcommand{\set}[3]{\{#1_{#2}\}_{#2=1}^{#3}}

\newcommand{\As}{\vv A_{s}}
\newcommand{\as}{\vv a_{s}}
\newcommand{\asi}{\vv a_{s}^{(i)}}

\newcommand{\Zs}{\vv Z_{s}}
\newcommand{\zs}{\vv z_{s}}
\newcommand{\deltafct}[1]{\delta\left[#1\right]}

\newcommand{\uu}{1\hspace{-4pt}1}
\newcommand{\fsur}{\vv f}
\newcommand{\avgc}[2]{\avg{#1}_{#2}}
\newcommand{\Cov}[1]{{\cal C}(#1)}
\newcommand{\Covc}[2]{{\cal C}(#1\mid #2)}
\newcommand{\zx}{z^{(x)}}
\newcommand{\zsx}{\zs^{(x)}}
\newcommand{\zsxi}{\zs^{(x,i)}}
\newcommand{\vzx}{\vv z^{(x)} }
\newcommand{\vcx}{\vv c^{(x)} }
\newcommand{\zsur}{z_\text{sur}}
\newcommand{\vzsur}{\vv z_\text{sur}}
\newcommand{\zsurx}{\zsur^{(x)}}
\newcommand{\vzsurx}{\vzsur^{(x)}}

\newcommand{\zxp}{z^{(x')}}
\newcommand{\vcxp}{\vv c^{(x')} }

\newcommand{\Ms}{M_{s}^{\phantom{T} } }

\newcommand{\MsT}{M_{s}^{T}}
\newcommand{\tr}[1]{\text{tr}\left\{ #1 \right \}}

\newcommand{\canceled}[1]{{\cancel{#1}}} 
We start with the general structure of uncertainty propagation problems based on surrogate models in Sec. \ref{sec:bayes_basics}.  In Sec. \ref{sec:surrogate_analysis}, we analyze  a generalized linear surrogate model with a Gaussian likelihood for the simulation data with unknown surrogate variance. In Sec. \ref{sec:UP} we proceed to use the surrogate model to propagate the input uncertainties to the output, and show how the surrogate's uncertainties too can be included.

\subsection{General structure of the problem} \label{sec:bayes_basics}

The goal in this paper is to quantify the uncertainties of the simulation results for the observable $\zx$ at different measurement points $x = 1,...,N_x$ in the simulation domain. E.g., $z$ could be the mechanical stress resulting from a structural analysis with a finite element simulation, where $x$ could denote the location of the measurement probe in or on the analysed structure. $\zx$  depends on unknown or uncertain model parameters $\vv a = \{a_i\}_{i=1}^{N_a}$, which are generally inferred from experimental data  $ \dexp$.
Based on these data, Bayes theorem allows  to determine the posterior probability density function (pdf) for $\vv a$,
\begin{align}\label{eq:aux0}
\p{\vv a}{ \dexp,\BK}\;, 
\end{align}
where all background information on the experiment is subsumed in $\BK$. The implications of the background information $\BK$ will be discussed later. This pdf will be assumed to be (almost) arbitrary but given in the following considerations. It usually is the result of a statistical data analysis of the foregoing experiment. This experiment could be the measurement of some material property needed for the simulation, e.g. viscosity for a computational fluid dynamics simulation.
The uncertainty of the model parameters $\vv  a$ entails an uncertainty in the simulated observable $\zx$, and the latter is determined by the marginalization rule,
\begin{align}\label{eq:aux1}
\p{\zx}{\dexp,\BK} &= \int \p{\zx}{\vv a,\canceled{\dexp},\BK}
\p{\vv a}{ \dexp,\BK} dV_{\vv a}\;.
\end{align}
In the first pdf we have striked out $\dexp$ because the knowledge of $\vv a$ suffices to perform the simulation  to obtain $z^{(x)}$. If $\vv a$ consists of only one or two parameters ($N_p = 1,2$), then the numerical evaluation of the integral over the model parameters $\vv a$ will typically require a few dozens to hundreds of simulations. The uncertainty propagation is then done and no surrogate is needed. However this is the trivial case, and usually $\vv a$ will consist of way more parameters. Let's assume that $\vv a$ consists of, e.g., four parameters. That would imply the need for performing  simulations at least $10^{5}$ times, which is way too CPU expensive for most real problems.
This can be avoided if the  simulations are replaced by a surrogate model  that approximates the observable $z$
by a suitable parametrized surrogate function $\zsur = g(\vv a|\vv c)$, where $\vv c$ are  yet unknown parameters. 
The simulation may yield the observable $z^{(x)}$  at different sites $x$ in the domain, however $x$ could also denote the time-instance in non-static problems. Clearly, 
the parameters of the surrogate model will also depend on those positions. So we actually have
\begin{align}\label{eq:surrogate_def}
\zx \approx \zsurx &= g(\vv a| \vcx)\;.
\end{align}
The unknown parameters  will be inferred from a suitable training data. To this end,   simulations are performed 
for a finite set of model parameters 
$\As =\{\vv a^{(i)}_{s}\}_{i=1}^{N_{s}}$
and the corresponding observables $\Zs = \{z_s^{(x),(i)}\}_{i,x=1}^{N_{s},N_x}$ are computed and combined in 
$\Dsim=\{\As,\Zs\}$. The surrogate parameters $\vcx$ are then inferred from $\Dsim$, and the surrogate is so constructed. We now proceed to substitute the simulation for the surrogate, $\zx \to \zsurx$, in order to solve \eqref{eq:aux1} at a significantly reduced computational cost. This implies that the background information has changed. We will denote this as  $\tilde \BK$, suggesting that we take the observable $z$ entering  the integral in \eqref{eq:aux1}  from the surrogate model \eqref{eq:surrogate_def} rather than from the expensive  simulation.
More precisely, instead of \eqref{eq:aux1} we now need to consider
\begin{align}\label{eq:aux2}
\p{\zx}{\dexp, \Dsim,\tilde \BK} &=
 \int dV_{\vv a}\;\p{\zx}{\vv a,\Dsim,\canceled{\dexp},\tilde \BK}\;
 \p{\vv a}{\canceled{\Dsim},\dexp,\tilde\BK} \;.
\end{align}
As far as the (second) pdf for the model parameters is concerned,  we can omit the information on the training set $\Dsim$, as it does not tell us anything about the model parameters.
This pdf is actually the same as that in \eqref{eq:aux0}, i.e. 
$\p{\vv a}{ \dexp,\tilde\BK}=\p{\vv a}{ \dexp,\BK}$,
 as it makes no difference 
for the model parameters how we solve the equations underlying the simulation.
In the first pdf, we can  omit the information on the experiment $\dexp$, as we only need the simulation data $\Dsim$ to fix the surrogate model, which in turn defines the observable $z$. 
The first pdf can be further specified by the marginalization rule upon introducing the surrogate  parameters $\vv C = \{\vcx\}_x$
\begin{align*}
 \p{\zx}{\vv a ,\Dsim  ,\tilde\BK}
 &=
\int dV_{\vv C} \;\p{\zx}{\vv C,\vv a ,\canceled{ \Dsim  },\tilde\BK}
\,
\p{\vv C}{\canceled{\vv a }, \Dsim,\tilde\BK   }\;.
\end{align*}
The first pdf is uniquely fixed by the knowledge of $\vv a$ and $\vv C$, hence $\Dsim$ is superfluous.
Similarly in the second pdf, where $\vv C$ is inferred from the training data, additional model parameters without the corresponding observables' values $z$, are useless.
In summary we have
\begin{align}\label{eq:aux3a}
\p{\zx}{\dexp,\Dsim,\tilde\BK } &=
 \iint  dV_{\vv a}
 dV_{\vv C}\; \p{\zx}{\vv C,\vv a,\tilde \BK }
\;
 \p{\vv C}{\Dsim ,\tilde\BK}\;
\p{\vv a}{ \dexp,\BK}
\;.
\end{align}
The first pdf is rather simple. According to the background information $\tilde \BK$ we will determine the observable via the surrogate model.
Since the necessary parameters $\vv c^{(x)}\in \vv C$  and $\vv a$ are part of the conditional complex, the surrogate model allows only one value
\begin{align} 
\zsurx &=  g( \vv a\mid \vcx)\;. \notag
\end{align}
for the observable. That means $\p{\vv z^{(x)}}{\vv C,\vv a,\tilde\BK } $ is equivalent to the probability density function for $\vv z^{(x)}$ given $\vv z^{(x)} =
g(\vv a\mid \vv c^{(x)})$.
Hence the pdf is a Dirac-delta distribution
\begin{align}\label{eq:dirac_delta}
\p{\zx}{\vv C,\vv a,\tilde\BK } 
&= \deltafct{\zx - g(\vv a\mid \vcx)}\;.
\end{align}
Finally, we have
\begin{align}\label{eq:aux3}
\p{\zx}{\dexp,\Dsim,\tilde\BK } &=
 \iint  dV_{\vv a}
 dV_{\vv C}\; \deltafct{\zx - g(\vv a\mid \vcx )}\;
\;
 \p{\vv C}{\Dsim ,\tilde\BK}\;
\p{\vv a}{ \dexp,\BK}
\;.
\end{align}
Before we can evaluate this integral, we first need to determine the 
two ingredients, which have their own independent significance. The last term is the result of a data analysis of a specific foregoing experiment, and will therefore not be treated here.
We will suppress the background information 
in the following, as ambiguities should no longer occur.

\subsection{Bayesian Analysis and Selection of the Surrogate Model}
\label{sec:surrogate_analysis}
We recall that \eqref{eq:aux3}  allows to determine the pdf for the
observable based on the pdf  for the model parameters, 
and the pdf for the parameters of the surrogate model $ \p{\vv C}{\Dsim}$, that we will determine now.
To this end we have to specify the form of the surrogate model.
We use the expansion
\begin{align} \label{eq:PCE_def}
\zsur  =  \sum_{\nu=1}^{N_{p}} \hparams _{\nu}   \Phi_{\nu}(\a)\;,
\end{align}
in terms of basis functions   $\Phi_{\nu}(\a)$ and expansion coefficients $c_{\nu}$. No further specification is needed at this point. Without loss of generality, we will use  multi-variate Legendre polynomials for the numerical examples. This expansion is very similar to the frequently used generalized Polynomial Chaos Expansion \citep{Xiu2005}, where the polynomials $\Phi_{\nu}(\a)$  are orthogonal with respect to the $L^2$ inner product with the prior of $\vv a$, $\prior{\a}$, as integration measure.
However, here we actually consider a posterior $\p{\a}{\de}$ that generally has no standard form, for which no standard orthogonal polynomial basis is known, and for which conditional independence of the model parameters $\vv a$ does not hold. The procedure has been extented to arbitrary probability measures \citep{Oladyshkin2012} and dependent parameters \citep{JAKEMAN2019}, but in the present context, however, these polynomials are not of primary interest and would only complicate the numerical evaluation.
As outlined in section \ref{sec:bayes_basics}, $N_s$  simulations are 
performed for a set of model parameters $\As =\{\vv a^{(i)}_{s}\}_{i=1}^{N_{s}}$ and the corresponding observables $\Zs$ are computed. The theory is so far agnostic to the experimental design of these simulations, and it is therefore not of concern here.
Now, we want to determine the pdf for the surrogate parameters $\vv C$, which are combined in a matrix with elements $C_{\nu,x}$, where $\nu$ enumerates the surrogate basis functions and $x$ enumerates the measurement positions in the domain for which the observables are computed.
We abbreviated the simulation data by the quantity $\Dsim=\{\As,\Zs\}$, where the matrix  $\Zs$ has the  elements    $(\Zs)_{i,x}$, which  represent the observable $\zx$ 
at position $x$  corresponding to the model parameter vector $\as^{(i)}$. 
The sought-for pdf follows from Bayes' theorem
\begin{align} 
\p{\vv C}{\Dsim} &=\p{\vv C}{\Zs, \As}
\propto \p{\Zs}{\vv C, \As} \p{\vv C}{\As}
\propto \p{\Zs}{\vv C, \As}. \notag
\end{align}
The proportionality constant is not required in the ensuing considerations
and we have  assumed an ignorant, uniform prior for the coefficients $\vv c$, i.e. $\p{\vv C}{\BK} = const$. We note that this is also the transformation invariant Riemann prior (see Appendix \ref{app:prior}).  However, any prior that is conjugate to the likelihood will retain analytical tractability.
For the likelihood we need the total misfit, which is given by
\begin{align}\label{eq:chi-min}
\chi^{2} &= \sum_{i=1}^{N_{s}}\sum_{x=1}^{N_{x}} \bigg(
(\Zs)_{i,x} - \sum_{\nu=1}^{N_{p}}   (M_{s})_{i,\nu} (\vv C)_{\nu,x}
\bigg)^{2}
=  \sum_{i,x} \bigg( \Zs - M_{s} \vv C\bigg)^2_{i,x}\nonumber\\
&= \tr{ \big( \Zs - M_{s} \vv C\big)^{T}  \big( \Zs - M_{s} \vv C\big)}
\end{align}
with $(M_{s})_{i,\nu} =     \Phi_{\nu}(\asi)$ and  $N_{sx} = N_{s}\cdot N_{x}$. We assume a Gaussian type of likelihood, i.e.
\begin{align}\label{eq:pdf:C}
    \p{\Zs} {\vv C,\As,\Delta}&=\frac{\Delta^{-N_{xs}}}{Z} \exp\left\{
    -\frac{ \chi^{2}}{2 \Delta^{2}}
    \right\}\;.
\end{align}
with normalization $Z$. 
We have mentioned the $\Delta$-dependence of the normalization explicitly, while the rest of of the normalization is irrelevant in the present context. 
Usually, the misfit entering the likelihood comes from the noise of the data. In the present case, however, there is no noise (merely a tiny numerical error), but 
the surrogate model is presumably not an exact 
description of the simulation data and $\Delta$ covers the corresponding uncertainty. However, the uncertainty level $\Delta$ is not known and 
has to be marginalized over. Along with the appropriate Jeffreys' prior, $\prior{\vv C, \Delta}{ }= \prior{\vv C}{ }\prior{\Delta}{ }$, $\prior{\Delta}{ }=\frac{1}{\Delta},\;\;\;\prior{\vv C}{ } = const.$ (see Appendix \ref{app:prior}), the integration over $\Delta$ yields 
\begin{align}\label{eg:marg:gauss:C}
    \p{\vv C} {\Zs,\As} &=\frac{1}{Z'} \big( \chi^{2} \big)^{-\frac{N_{sx}}{2}}\;.
\end{align}
with terms independent of $\vv C$ subsumed in the normalization $Z'$. 
  {
For computing the mean, variance and evidence we first complete the square in \eqref{eq:chi-min} to get a quadratic form in $\vv C$, which can then be integrated analytically (see Appendix \ref {app:proofs}). The result is
}
\begin{subequations}  \label{eq:C-moments}
\begin{align} \label{eq:C_mean}
\avgc{\vv C}{\vv a} &= H_{s}^{-1} M_{s}^{T}\Zs \;, &
H_s &= M_s^T M_s  \;,
\\ \label{eq:C_covar}
\avgc{\Delta C_{\nu x} \Delta C_{\nu' x'}}{\vv a} 
&= \frac{ \chi^{2}_\text{min} }{
(N_{s}-N_{p})N_{x}-2}\; \big(H_{s}^{-1}\big)_{\nu,\nu'} \;\delta_{xx'}\;, &
\chi^{2}_\text{min}&= \tr{
\Zs^{T}
\big( \uu - M_{s} H^{-1}_{s} M_{s}^{T} \big)
\Zs}\;.
\end{align}
\end{subequations}
%
and we argue that the prefactor of $H_{s}^{-1}$ is the Bayesian estimate for $\Delta^{2}$, the variance of the Gaussian in \eqref{eq:pdf:C}.   {This reasoning is similar as in \citep{VDL1996}}. Note that $\Zs$ is a matrix of size $N_s \times N_x$, containing the data vectors of length $N_s$ for each measurement site $x$. As shown in Appendix \ref {app:proofs}, the evidence for a particular set of surrogate models is computed as
\begin{align}\label{eq:evidence}
\p{\{\zsurx\}_{x=1}^{N_x}}{\Dsim, \tilde{\BK}} = Z'
= \Omega_{N_{bx}}\;{\lvert H_s\rvert}^{-\frac{1}{2}}
\big( \chi^{2}_\text{min} \big)^{- \frac{N_{sx}-N_{bx}}{2}}\; \frac{
\Gamma(\frac{N_{bx}}{2})
\Gamma(\frac{N_{sx}-N_{bx}}{2})
}{\Gamma(\frac{N_{sx}}{2})}
\end{align}
where $\Omega_{N_{bx}}$ is the solid angle in $N_{bx}$ dimensions. The evidence is the probability for a surrogate model given the data. Note that this quantity does not depend on $\p{\vv a}{\dexp, \BK}$. This is reasonable because the analysis of the experimental data should be independent of the analysis of the simulation data. However $\p{\vv a}{\dexp, \BK}$ will typically be used for the experimental design of the simulation data acquisition. By comparing the evidences for different models, the user can choose a particular surrogate model or, if the results do not overwhelmingly suggest one single model, average the results for the surrogate analysis and the following uncertainty propagation over several plausible models. Note that the evidence is the pillar of a Bayesian procedure to select a surrogate model, and is distinct to the procedure of incorporating the trustworthiness or uncertainty of the surrogate in the subsequent uncertainty propagation.

%

\subsection{Bayesian Uncertainty Propagation with Surrogate Models} \label{sec:UP}

Now that we have selected the surrogate model and determined the ingredients of \eqref{eq:aux3},
we can determine the pdf for the observables in the light of the experimental data and the simulation results of the training set.
The form in \eqref{eq:aux3a} allows an easy evaluation of the mean value
\begin{align}\label{eq:z:avg}
\avg{\zx} &= 
 \iint  dV_{\vv a}
 dV_{\vv C}\; f(\vv a\mid \vcx)
\;
 \p{\vv C}{\Dsim,\tilde\BK}\;
\p{\vv a}{ \dexp,\BK}\nonumber\\
&=\sum_{\nu} 
 \int  dV_{\vv a} 
\Phi_{\nu}(\vv a)  \avgc{C_{\nu x}}{\vv a}
\;
\p{\vv a}{ \dexp,\BK}
\end{align}
Similarly we obtain 
\begin{align}
\label{eq:z:cov}
\avg{\zx \zxp} &= 
 \iint  dV_{\vv a}
 dV_{\vv C}\;
\;f(\vv a\mid \vcx)
\;
f(\vv a\mid \vcxp)  
\;
\;
 \p{\vv C}{\Dsim ,\tilde\BK}\;
\p{\vv a}{ \dexp,\BK}\nonumber\\
&= 
\sum_{\nu\nu'}
 \int  dV_{\vv a}
\;\Phi_{\nu}(\vv a)
\;\Phi_{\nu'}(\vv a)
\;
\avgc{C_{\nu x} C_{\nu' x'}}{\vv a}
\;
\;
\p{\vv a}{ \dexp,\BK}\nonumber\\
&= 
\sum_{\nu\nu'}
 \int  dV_{\vv a}
\;\Phi_{\nu}(\vv a)
\;\Phi_{\nu'}(\vv a)
\;
\bigg(
\avgc{C_{\nu x}}{\vv a} \avgc{C_{\nu' x'}}{\vv a}
+
\avgc{\Delta C_{\nu x} \Delta C_{\nu' x'}}{\vv a} 
\bigg)
\;
\;
\p{\vv a}{ \dexp,\BK}\;.
\end{align}
The covariance then follows from
\begin{align} 
\avg{\Delta z^{(x)} \Delta z^{(x')}} &= 
\avg{z^{(x)}z^{(x')}} -\avg{z^{(x)}}\avg{ z^{(x')}}\;.
\end{align}
If we neglected the uncertainty of the surrogate, i.e. 
\begin{align}
    \p{C}{D_{sim}} &= \delta (C -\hat C) \notag \\
    \hat C = \avgc{C}{\vv a}  \notag
\end{align}
then we retain the widely known special case of 'perfectly trustworthy' surrogates
\begin{align}
\avg{\zx \zxp}  &=  \sum_{\nu\nu'}  \int  dV_{\vv a} \;\Phi_{\nu}(\vv a) \;\Phi_{\nu'}(\vv a)
\; \avgc{C_{\nu x}}{\vv a} \avgc{C_{\nu' x'}}{\vv a}
\;
\;
\p{\vv a}{ \dexp,\BK}\;. \notag
\end{align}
Thus, the first part in the integral of \eqref{eq:z:cov} is the uncertainty of the observable due to experimental uncertainties and given the surrogate model, while the second term adds the uncertainty of the surrogate itself. The term $\avgc{\Delta C_{\nu x} \Delta C_{\nu' x'}}{\vv a}$ is commonly neglected, but easily computed. This result also suggests a natural measure for the trustworthiness of the surrogate model, which is directly linked to the specific experiment:
\begin{align}
    \frac{\sum_{\nu\nu'}  \int  dV_{\vv a} \;\Phi_{\nu}(\vv a) \;\Phi_{\nu'}(\vv a)
\; \avgc{\Delta C_{\nu x} \Delta C_{\nu' x'}}{\vv a} \;\;
\p{\vv a}{ \dexp,\BK}\;}{\sum_{\nu\nu'}  \int  dV_{\vv a} \;\Phi_{\nu}(\vv a) \;\Phi_{\nu'}(\vv a)
\; \avgc{C_{\nu x}}{\vv a} \avgc{C_{\nu' x'}}{\vv a}
\;
\;
\p{\vv a}{ \dexp,\BK}\;} < \epsilon
\end{align}
If the surrogate uncertainties are, on average, smaller than the experimental uncertainties by a few orders of magnitude, e.g. $\epsilon=10^{-3}$, then they may be neglected. However $\epsilon$ is the user's choice. Note that this result does not spare the user to solve the foregoing surrogate model selection problem by e.g. computing evidences. This work only demonstrates how surrogate uncertainties can be incorporated and a practical rule when they could be neglected, given the surrogate model has already been selected before.

\section{Numerical Example}
Here, we demonstrate an application where surrogate uncertainties were in part negligible and in part non-negligible. 
We apply our method to a computational fluid dynamics simulation of aortic hemodynamics, i.e. blood flow in an aorta resembled by the simplified geometry of an upside down umbrella stick. The simulation depends on a non-Newtonian viscosity model with four parameters $\vv a = \{a_1, a_2, a_3, a_4\}$. The model was accompanied by viscosity measurements of human blood samples, thus determining $\p{\vv a}{ \dexp,\BK}$. This posterior turned out to have a complex landscape that cannot be reasonably approximated by standard distributions. Particularly, strong correlations and multi-modality was observed, i.e. $\p{\vv a}{ \dexp,\BK} \neq \prod_i \p{a_i}{\dexp, \BK}$. This means, that vanilla Polynomial Chaos Expansions could not be applied without an undesirable transformation to conditionally independent variables. The posterior is described in detail in \citep{Ranftl2021a_dataset}. Based on $\p{\vv a}{ \dexp,\BK}$, $N_s = 100$ parameter samples $\vv a_s$ were chosen and the simulation evaluated accordingly. The output, $\Zs$, were the absolute values of the wall shear stress, that the blood flow exerts on the aortic wall, for $N_x = 10$ measurement probes at different locations, each for $N_t=101$ time-instances equidistantly spaced over one cardiac cycle (ca. 1 sec). Further details on the simulation are not relevant here, but are documented  \citep{Ranftl2021a_dataset}. A simulation time on the order of $150$ CPU hours per sample suggested to use a surrogate for the inference. For the surrogate's basis functions, $\Phi_{\nu}(\vv a)$, we found multi-variate Legendre polynomials up to order two sufficient. The numerical integrals were computed with Riemannian quadrature and convergence checked with successive grid refinement, however stochastic integration would work just as well. A sketch example on how to implement this procedure computationally efficient via vectorisation in parameter space can be found at \url{https://github.com/Sranf/BayesianSurrogate_sketch.git}
 
\begin{figure}
    \centering
    \includegraphics[width=\textwidth]{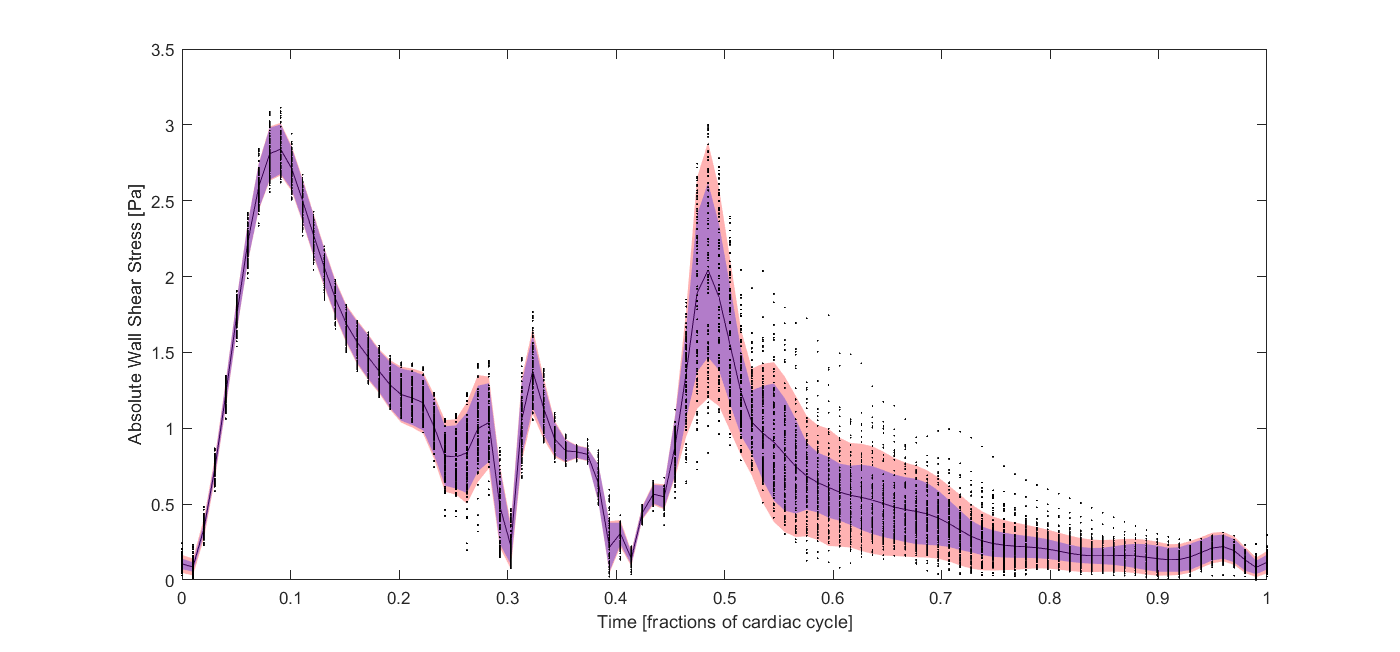}
    \caption{Simulation data (black dots) and simulation uncertainty ($1\sigma$) according to our Bayesian approach (red, including surrogate uncertainty) as well as the naive simulation uncertainty (blue, neglecting surrogate uncertainty). The black line is the surrogate mean.}
    \label{fig:numerical-example}
\end{figure}

In \figref{fig:numerical-example}, we compare the simulation uncertainty (including surrogate uncertainty) as computed with our Bayesian approach (\eqref{eq:z:cov}) to the naive estimate for the simulation uncertainty (without surrogate uncertainty, i.e. neglecting $\avgc{\Delta C_{\nu x} \Delta C_{\nu' x'}}{\vv a}$ in \eqref{eq:z:cov}). The surrogate uncertainties in the first half (left hand side) are relatively small, comprising only a few percent of the total uncertainty, and could possibly be neglected. In the second half (right hand side) however, the surrogate uncertainties make up to $\sim 50\%$ of the total uncertainty. This demonstrates that simulation uncertainties inferred via surrogate models can be severely underestimated if the surrogate uncertainties are neglected, and subsequently lead to overconfidence in the simulation model. In practise, one would acquire more data in order to reduce the surrogate uncertainties, e.g. more data at later time-instances in \figref{fig:numerical-example} is particularly promising. This was here limited not only by the computational budget, but also unpractical in that dynamic simulations require the full evaluation of all previous time instances where the surrogate is already reasonably accurate. Thus, the procedure of instead explicitly including the surrogate uncertainties here also has proven to be practical. A similar situation is to be expected for most transient simulations, as uncertainties will usually increase as time progresses.
 
\section{Discussion} \label{sec:conclusion}
In this work, we have assumed a Gaussian likelihood for the simulation data, with unknown variance, for a surrogate that is linear in its parameters. Surrogates that are non-linear in its parameters (e.g. neural networks) may promise higher capacity, however at the expense of losing analytical tractability of the surrogate uncertainty entirely. Other likelihood functions might be useful if further information is available, such as bounds on the observable (Gamma- or Beta-likelihood).

The result is a simple formula to incorporate surrogate uncertainties in the simulation uncertainties. This formula will be particularly useful if 'convergence' in the sense of finding the coefficients of e.g. a Polynomial Chaos Expansion is doubtful or not achievable due to a limit to the computational budget.  The formula immediately suggests an intrinsic measure for the trustworthiness of the surrogate, distinct from commonly used ad-hoc diagnostics. This measure is not to be confused with the evidence, and should not be used for model selection because it would not preclude over-fitting etc. It is merely a measure for the trustworthiness of the already selected surrogate.

Let us now explore the connections of this work to Polynomial Chaos Expansions (PCE) and Gaussian Process Regression (GPR). PCE is a special case of our generalized linear surrogate model, in that the basis functions of the surrogate are chosen such that 
\begin{align} \label{eq:pce-def-nb}
    \int \Phi_{\nu} (\vv a) \Phi_{\nu'}(\vv a) \p{\vv a}{\dexp, \BK} dV_{\vv a} := \delta_{\nu,\nu'} \;.
\end{align}
The double sum in \eqref{eq:z:cov} then contracts to a single sum, and the diagonal of the term for the surrogate uncertainty, $\avgc{\Delta C_{\nu x} \Delta C_{\nu' x'}}{\vv a}$, survives. This is expected, in that PCE is defined such that the basis functions are uncorrelated, but still the expansion coefficients must be uncertain to a finite degree and this must carry over to the simulation uncertainty. 
A severe limitation of PCE is, that it is rather difficult to find basis function sets $\{\Phi_{\nu}\}$ that fulfill \eqref{eq:pce-def-nb}, depending on $\p{\vv a}{ \dexp,\BK}$. For most practical purposes, one demands (i) conditional independence of the simulation parameters, i.e. $\p{\vv a}{ \dexp,\BK} \neq \prod_i \p{a_i}{\dexp, \BK}$, as well as (ii) simple standard distributions for $\p{a_i}{\dexp, \BK}$, in order to find a solution (usually a tensor-product) to \eqref{eq:pce-def-nb}. Known albeit tedious work-arounds are for (i) variable transformations and numerical orthonormalisation \citep{JAKEMAN2019} and for (ii) PCE-constructions for arbitrary pdfs \citep{Oladyshkin2012}. Note that also \citep{Oladyshkin2012} demands (i) conditional independence of the simulation parameters.
In the numerical example above, neither (i) nor (ii) were applicable. Finding a variable transformation in order to fulfill (i) or numerical construction of orthonormal basis functions can be difficult, and particularly inconvenient if sophisticated priors $\p{\vv a}{\BK}$ are being used, e.g. Jeffreys' generalized prior.   {An interesting alternative would be to model the input dependencies with vine copulas \citep{TORRE2019} in order to overcome the limitations of PCE addressed here. Unfortunately, no obvious vine copula was found for the here presented example.}

Gaussian Process Regression would correspond to a change
in the prior for $\zx$ in \eqref{eq:dirac_delta} as follows
\begin{align}
   \p{\zx}{\vv C,\vv a,  {\theta}, \tilde\BK } 
&= \mathcal{N}\Big(g(\vv a\mid \vcx) \Big\lvert  K(\vv \theta) \Big)\;. 
\end{align}
where $\mathcal{N}$ denotes a normal distribution and $K$ is the prior's covariance matrix and defined by the parametrized covariance function $k$ , $[K]_{ij} = k(\vv a^{(i)}, \vv a^{(j)} \mid \vv \theta)$. This in turn would change $(\Zs - M_s \vv C)^T (\Zs - M_s \vv C) \to (\Zs - M_s \vv C)^T K^{-1} (\Zs - M_s \vv C)$ in \eqref{eq:chi-min}.   {By again completing the square and following the same procedure,} the corresponding results for mean \eqref{eq:C_mean}, variance \eqref{eq:C_covar} and evidence \eqref{eq:evidence}  are then retained by a simple substitution of
\begin{align*}
    H_s &\to \tilde H_s  & \quad  
    \tilde H_s &= M_s^T K_s^{-1} M_s \\
    \chi_{min}^2 &\to \tilde{\chi}_{min}^2 & \quad  
    \tilde{\chi}_{min}^2 &= \tr{ \Zs^T (K_s^{-1} - K_s^{-1} M_s \tilde{H}_s M_s^T K_s^{-1}) \Zs }
\end{align*}
where  $[K_s]_{ij} = k(\as^{(i)}, \as^{(j)} \mid \vv \theta)$ is the likelihood's covariance matrix evaluated  at $\As$ for the data set $\Zs$ at given $\vv \theta$. \eqref{eq:z:avg} and \eqref{eq:z:cov} would preserve their form with the substitution
\begin{align*}
    \Phi_{\nu}(\vv a) \avgc{\vv C}{\vv a} & \to  \Phi_{\nu}(\vv a) \avgc{\vv C}{\vv a, \vv \theta} + K_{\ast}^T K_s^{-1} M_s \avgc{\vv C}{\vv a, \vv \theta} \\
    \avgc{\Delta C_{\nu x} \Delta C_{\nu' x'}}{\vv a} &\to
    \avgc{\Delta C_{\nu x} \Delta C_{\nu' x'}}{\vv a, \vv \theta} \Big(K -  K_{\ast}^T K_s^{-1} K_{\ast} \Big) \;,
\end{align*}
  {where the subscript $\theta$ acknowledges that the right hand side now depends on $\vv \theta$ and $[K_{\ast}]_{ij} = k(\vv a^{(i)}, \as^{(j)} \mid \vv \theta)$ is the covariance between the training set $\As$ and the 'test set', i.e. the integration variable $\vv a$.
}
Note that the additionally introduced hyperparameters $\vv \theta$ would require the choice of a prior for $\vv \theta$ and marginalization wrt $\vv \theta$ in \eqref{eq:aux3}, and subsequently also (\ref{eq:C-moments}- \ref{eq:z:cov}). 

We now discuss the implications of $\tilde{ \mathcal{I}}$ in contrast to the original background information $\BK$ .
$\mathcal{I}$ contains, most importantly, that the observable $z$ is uniquely determined by the simulation for a given set of input parameters $\vv a$. A prerequisite here was, that the simulation is converged. E.g. for finite element simulations this would be a given mesh-converged spatial discretization. The proposition $\tilde{ \mathcal{I}}$ additionally assumes \eqref{eq:surrogate_def} and \eqref{eq:dirac_delta}, so that it can be used to replace \eqref{eq:aux1} by \eqref{eq:aux2}. Formally this means, to get from  \eqref{eq:aux1} to \eqref{eq:aux2}, we replace $p(z\mid \vv{a}, \mathcal{I}) \to p(z\mid \vv{a}, \vv{c}, \tilde{\mathcal{I}})$, where 
$p(z\mid \vv{a}, \mathcal{I}) = \delta (z - z(\vv{a}) )$, $p(z\mid \vv{a}, \vv{c}, \tilde{\mathcal{I}}) = \delta (z - z_{sur}(\vv{a}) ) = \delta (z - g(\vv{a}\mid \vv{c}) )$. 
I.e. $\tilde{ \mathcal{I}}$ contains in comparison to $\mathcal{I}$ the additional assumption that we can use the value for $z$ as predicted/approximated by the surrogate model. It also means that we introduce additional, artificial, and usually unknown regression parameters $\vv{c}$ that need to be marginalized over. The additional uncertainty introduced by this approximation (i.e. the surrogate assumption) is encoded in $p(\vv{c} |\Dsim, \tilde{ \mathcal{I}})$, and is correctly incorporated into the simulation observable uncertainties in \eqref{eq:z:cov}.
What is important here, is that $p(z \mid \dexp, \Dsim, \tilde{\mathcal{I}}) \neq p(z \mid \dexp, \cancel{\Dsim}, {\mathcal{I}} ) $ in general (the latter is computationally infeasible) but $p(z \mid \dexp, \Dsim, \tilde{\mathcal{I}}) \approx p(z \mid \dexp, \Dsim, {\mathcal{I}} ) $  if \eqref{eq:surrogate_def} holds and  $p(\vv{C}\mid \Dsim) \approx \delta (\vv{C} - \hat{\vv{C}})$, i.e. if the surrogate is indeed a good approximation and the posterior for the surrogate parameters is sharply peaked at $\hat{\vv{C}}$. Very often this posterior is not sharply peaked, then we can just gather more data until it is, or, if that is not possible, we can at least avoid overconfidence induced by neglecting these uncertainties. A numerical example of where this is the case has been demonstrated above. 

We have modelled spatial correlations by introducing a location index $x$, and assumed that the expansion coefficients at different sites, $\vv c ^{(x)}$ and $\vv c ^{(x')}$, are conditionally independent. This assumption is reasonable, in that the expansion coefficients are arbitrary mathematical constructs and no physically motivated model for their correlation is known. The spatial correlation however is retained in $\zx$, as was originally intended.  More general models for spatial correlations can easily be implemented by substitution of $\delta_{xx'}$ with a spatial covariance matrix in \eqref{eq:C_covar}. Note, that this would require an additional marginalization wrt the (typically non-linear) hyperparameters of the spatial covariance matrix.
By introducing a compound index $\tilde x = (x, t)$ and substituting $x\to \tilde x$, we find a simple generalization to spatio-temporal correlations. This is equivalent to re-ordering spatial and temporal indices into a single sequence. While this  procedure is convenient and requires only minor changes in the numerical implementation, it implicitly assumes conditional independence of spatial and temporal correlations. Analogous to above, general temporal correlations can be modelled by a substitution of $\delta_{\tilde x \tilde x '}$ in \eqref{eq:C_covar} with a temporal covariance matrix, again requiring an additional marginalization wrt the latter's hyperparameters.

\section{Conclusions}

We presented a Bayesian analysis of surrogate models and its associated uncertainty propagation problem in the context of uncertainty quantification of computer simulations. The assumptions were a generalized linear surrogate model (linear in its parameters, not the variable) and a Gaussian likelihood with unknown variance. Additionally, spatial and temporal correlations have been discussed.
The result suggests a measure of trustworthiness of the surrogate by quantifying the ratio of the surrogate uncertainty to the total uncertainty, in contrast to commonly used heuristic diagnostics. The main result however is a rather simple rule to include surrogate uncertainties in the sought-for uncertainties of the simulation output. This is useful particularly for problems where the surrogate's trustworthiness is doubtful and cannot be improved. The connections to Polynomial Chaos Expansions and Gaussian Process Regression have been discussed. A numerical example demonstrated that simulation uncertainties can be significantly underestimated if surrogate uncertainties are neglected.


\vspace{6pt}

\section*{Funding}{This work was funded by Graz University of Technology (TUG) through the LEAD Project "Mechanics, Modeling, and Simulation of Aortic Dissection" (biomechaorta.tugraz.at) and  supported by GCCE: Graz Center of Computational Engineering.}



\section*{Data availability}{All information is contained in the manuscript. Code sketches are available here: \url{https://github.com/Sranf/BayesianSurrogate_sketch.git}
} 

\section*{Acknowledgments}{The authors are grateful for useful comments from Ali Mohammad-Djafari.}

\section*{Conflicts of interest}{The authors declare no conflict of interest.} 







\bibliographystyle{unsrtnat}
\bibliography{references}

\appendix
\section{Mathematical proofs} \label{app:proofs}

\newcommand{\abs}[1]{\left| #1 \right|}

Here we want to determine norm, mean and covariance of the
 marginalized Gaussian (Student-t distribution) in \eqref{eg:marg:gauss:C},
 which is
\begin{align}
\p{\vv C}{\Zs,\As} &= \frac{1}{Z'} \big( \chi^{2} \big)^{-\frac{N_{sx}}{2}}\;, \nonumber\\
\chi^{2} &= \tr{\big( \Zs - M_{s} \vv C \big)^{T}  \big( \Zs - M_{s} \vv C \big) }\;.
\end{align}
  {
In order to perform the integration, we first complete the square to get a quadratic form in $\vv C$, which can then be integrated analytically,
}
i.e. we bring  the misfit $\chi^2$ into a form that elucidates the $\vv C$ dependence 
\begin{subequations} 
\begin{align}
\chi^{2}
&= \chi^{2}_\text{min} + \tr{\big(\vv C-\hat{\vv C}\big)^{T} H_s \big(\vv C-\hat{\vv C}\big)}\;, & 
H_{s} &= M_{s}^{T} M_{s}\;,\\
\chi^{2}_\text{min}&= \tr{
\Zs^{T}
\big( \uu - M_{s} H^{-1}_{s} M_{s}^{T} \big)
\Zs}\;,  & \hat{\vv C} &= H_{s}^{-1} M_{s}^{T} \Zs\;. 
\end{align}
\end{subequations}
%
Now, the first moment is easily obtained.
Along with the variable transformation under the integral 
\begin{align}\label{eq:var:transform}
\vv C &\to \hat{ \vv C} + \vv X
\end{align}
we obtain
\begin{align}
\avg{\vv C} &=\frac{1}{Z} \int dV_{\vv C}\; \vv C\;\bigg ( \tr{ (\vv C- \hat {\vv C})^T H_s (\vv C- \hat {\vv C})} + \chi^{2}_\text{min}\bigg)^{-\frac{N_{sx}}{2}}
\;\nonumber\\
&=\hat{\vv C} + \frac{1}{Z}\underbrace{
 \int dV_{\vv X} \; \vv X\; \bigg(  \vv X^T H_s  \vv X+ 
\chi^{2}_\text{min} \bigg)^{-\frac{N_{sx}}{2}} 
}_{  = 0} \;.
\end{align}
where we have used the symmetry properties of the likelihood.
Next, we transform the expression for  normalization 
 based on \eqref{eq:var:transform}
\begin{align}
Z_{N_{sx}} &= \int dV_{\vv X}\;\bigg ( \tr{
 \vv X^{T} H_s \vv X} + \chi^{2}_\text{min}\bigg)^{-\frac{N_{sx}}{2}}\;.\nonumber
\end{align}
Now we combine and reorder the double indices $(\nu, x)$ into a single index $l$, which
turns the matrix $\vv X$ of dimension $N_p \times N_x$ into a 
vector $\vv x$ of dimension $N_{px} = N_{p}\cdot N_{x}$ and the matrix $H$ of dimension $N_p \times N_p$ into a new block matrix $\vv H$ of dimension $N_{px} \times N_{sx}$ such that
\begin{align} \label{eq:reordering}
 [\vv H ]_{ll'}   &= [H_s]_{\nu,\nu'} \; \delta_{xx'}
\end{align}
In this representation we have
\begin{equation}
Z_{N_{sx}} = \int dV_{\vv x}\;
\bigg ( 
 \vv x^{T} \vv H \vv x + \chi^{2}_\text{min}
 \bigg)^{-\frac{N_{sx}}{2}} = |\vv H|^{-\frac{1}{2}}\; \int dV_{\vv y}\;\bigg ( 
\vv y^{T}  \vv y + \chi^{2}_\text{min}\bigg)^{-\frac{N_{sx}}{2}}\;.
\end{equation}
where we substituted $\vv x \to \vv H^{-\frac{1}{2}} \vv y$.
Next we  introduce hyper-spherical coordinates, which leads to
\begin{align*}
Z_{N_{sx}} &=\Omega_{N_{bx}}\;\abs{\vv H}^{-\frac{1}{2}}\;\int_{0}^{\infty}	 \frac{d\rho}{\rho} \;\rho^{N_{bx}}  ( \rho^{2} + \chi^{2}_\text{min})^{-\frac{N_{sx}}{2}}\;,
\end{align*}
where $\Omega_{N_{bx}}$ is the solid angle in $N_{bx}$ dimensions. 
Finally, based on the  substitution $\rho = t \cdot \sqrt{\chi^{2}_\text{min}}$, we recover an identity of the Beta-function and we obtain
\begin{align}
Z_{N_{sx}} 
&=
\Omega_{N_{bx}}\;\abs{\vv H}^{-\frac{1}{2}}\;
\big( \chi^{2}_\text{min} \big)^{- \frac{N_{sx}-N_{bx}}{2}}\; \frac{
\Gamma(\frac{N_{bx}}{2})
\Gamma(\frac{N_{sx}-N_{bx}}{2})
}{\Gamma(\frac{N_{sx}}{2})}
\end{align}
This result is valid only for $N_{sx}>N_{bx}$, which is fulfilled in the present application. For future use we rewrite this as
\begin{align}
Z_{N_{sx}} 
&=Z_{(N_{sx}-2)}\;\cdot\;\big(\chi^{2}_\text{min}\big)^{-1} \;\cdot 
\frac{
N_{sx}-N_{bx}-2
}{N_{sx}-2}\;.
\end{align}
Finally, we calculate the covariance, based also on the compound index
$l = (\nu, x)$, and by using  the  variable transformation in \eqref{eq:var:transform}.
\begin{align}
\avg{\Delta C_{l} \Delta C_{l'}}  &=\frac{1}{Z_{N_{sx}}}\;  \int dV_{\vv x}\;x_{l}x_{l'}\;\bigg(\vv x^{T} \vv H \vv x+\chi_\text{min}^{2}\bigg)^{-\frac{N_{sx}}{2}}\;,\nonumber\\
&= -\frac{2}{N_{sx}-2}\cdot \frac{1}{Z_{N_{sx}}}\cdot \frac{\partial }{\partial \vv H_{l,l'}}  \int dV_{\vv x}\;\bigg(\;\vv x^{T} \vv H \vv x+\chi_\text{min}^{2}\bigg)^{-\frac{N_{sx}}{2}+1}\;,\nonumber\\
&= -\frac{2}{N_{sx}-2} \frac{\chi^{2}_\text{min} \cdot (N_{sx}-2)}{Z_{(N_{sx}-2)}\cdot(N_{sx}-N_{bx}-2)}\; \frac{\partial }{\partial \vv H_{ll'}}  Z_{(N_{sx}-2)}\;,\nonumber\\
&= - \frac{2 \chi^{2}_\text{min} }{(N_{sx}-N_{bx}-2)}\; \frac{\partial }{\partial \vv H_{ll'}}
\ln(Z_{(N_{sx}-2)})\;,\nonumber\\
&= - \frac{2 \chi^{2}_\text{min} }{(N_{sx}-N_{bx}-2)}\; \underbrace{\frac{\partial }{\partial \vv H_{ll'}}
\ln(|\vv H|^{-\frac{1}{2}})}_{=-\frac{1}{2}\big( \vv H^{-1}\big)_{ll'}}\;,\nonumber\\
\end{align}
In the last step we have used that  $\vv H$ is a symmetric matrix.
This is a very reasonable result because if the variance $\Delta^{2}$ in the Gaussian in \eqref{eq:pdf:C} would be known,
then the covariance is $\Delta^{2} \vv H^{-1}$. Consequently, the prefactor represents the Bayesian estimate for the variance $\Delta^{2}$ based on the data.
Now we go back to the original meaning of the compound index \eqref{eq:reordering}, i.e. $\big(\vv H^{-1}\big)_{ll'} \to  (H^{-1}_{s})_{\nu\nu'} \delta_{xx'}$, and obtain the final result.

\section{The transformation invariant prior for the surrogate coefficients} \label{app:prior}
Bayesian probability theory allows to rigorously and consistently incorporate any prior knowledge we have about the experiment before taking a look at the data. This knowledge shall be elicited here.
Our inference must not depend on the exact parametrization. E.g. if we re-parametrize the surrogate, re-label or re-order the surrogate parameters, the surrogate still should describe the same simulation. This is reasonable because the surrogate is a purely mathematical, auxiliary construct. This rescaling-invariance is ensured by Jeffreys' generalized prior and is given by the Riemann metric  $R$ (or the determinant of the Fisher information matrix) \citep{vonderLinden2014}
\begin{align}\label{eq:prior_def_riemann}
p(\vv C) &= \frac{1}{Z} \big| \det(R)\big|^{1/2} && \text{with} &
R_{ij} &= \int  \p{\vv \Zs}{\vv C} 
\frac{\partial^{2} }{\partial C_{i}\partial C_{j}}
\ln\big(
\p{\vv \Zs}{\vv C} \big) \; dV_{\Zs}\;.
\end{align}
with multi-indices $i,j = (\nu,x)$. 
With the likelihood and the generalized surrogate model defined in the manuscript, the result is
\begin{align}\label{eq:riemann_metric:b}
R_{ij} &\propto \sum_{k=1}^{N_s} 
\frac{\partial g(\vv a^{(k)}_{s} \mid \vv C)}{\partial C_{i}} \notag
\frac{\partial g(\vv a^{(k)}_{s} \mid \vv C)}{\partial C_{j}}\\ \notag
&= \sum_{k=1}^{N_s} \Phi_i(\vv a^{(k)}_{s}) \Phi_j(\vv a^{(k)}_{s})\\
&=  const.
\end{align}
This prior is independent of $\vv C$, i.e. a constant.

\end{document}